# Strongly coupled phase transition in ferroelectric/correlated electron oxide heterostructures

Lu Jiang,[1,2] Woo Seok Choi,[1] Hyoungjeen Jeen,[1] Takeshi Egami,[1,2] and Ho Nyung Lee[1,a]

[1]*Materials Science and Technology Division, Oak Ridge National Laboratory, Oak Ridge, TN 37831, USA*

[2]*Department of Physics and Astronomy, University of Tennessee, Knoxville, TN 37996, USA*

[a]Author to whom correspondence should be addressed. Electronic mail: hnlee@ornl.gov

**We fabricated ultrathin ferroelectric/correlated electron oxide heterostructures composed of the ferroelectric $Pb(Zr_{0.2}Ti_{0.8})O_3$ and the correlated electron oxide (CEO) $La_{0.8}Sr_{0.2}MnO_3$ on $SrTiO_3$ substrates by pulsed laser epitaxy. The hole accumulation in the ultrathin CEO layer was substantially modified by heterostructuring with the ferroelectric layer, resulting in an insulator-metal transition. In particular, our thickness dependent study showed that drastic changes in transport and magnetic properties were strongly coupled to the modulation of charge carriers by ferroelectric field effect, which was confined to the vicinity of the interface. Thus, our results provide crucial evidence that strong ferroelectric field effect control can be achieved in ultrathin (10 nm) heterostructures, yielding at least a 100,000-fold change in resistivity.**

Sr-doped LaMnO$_3$ (La$_{1-x}$Sr$_x$MnO$_3$, LSMO) has a rich phase diagram and exhibits a wide spectrum of magnetic and electronic properties, mainly due to the strong coupling among spin, charge, orbital, and lattice degrees of freedom.[1-3] The phase diagram of LSMO reveals numerous different phases with increasing $x$ (hole doping), as partially shown in Fig. 1(a): spin-canted insulator (C-I), ferromagnetic insulator (FM-I), ferromagnetic metal (FM-M), and antiferromagnetic insulator (AF-I). The complicated phase diagram manifests the material's sensitivity to external stimuli, such as thickness,[4-8] strain,[9-11] microstructure, and lattice distortion,[12] in addition to chemical doping and temperature. In particular, in the vicinity of its phase boundaries, for instance $x = \sim0.2$, the phase of LSMO is most susceptible to a small perturbation and, thus, even an insulator-metal transition (IMT) can be realized in controllable fashion.

As one of controlling parameters, electric polarization in ferroelectrics can be efficiently used to tune the characteristics of LSMO. In oxide heterostructures, it has been frequently shown that the physical property of LSMO could be manipulated by the neighboring ferroelectric layer.[13-17] Even though the concept of ferroelectric field effect seems promising, the direct probing of the field-induced change in electronic and magnetic ground states with a conventional capacitor as a function of the polarization direction has been rather limited due mainly to the following reason: It is practically impossible to switch the polarization across a large area of a macroscopic ferroelectric layer (typically several mm$^2$) needed for DC transport and magnetic measurements, since the required energy to switch the polarization is extremely high. Even if one could reach the required energy level, the typical size of the top electrode should be relatively small (in the order of 100 $\mu$m in diameter or less). Otherwise, when a larger capacitor is used, a huge charging current due to a large RC time constant would hinder the reliable ferroelectric switching. In addition, it is almost impossible to avoid unwanted pinholes or particulates in thin films prepared by most physical vapor depositions. To overcome this problem, some studies used sub-micron-size field effect transistor and observed ferroelectric switching dependent changes in the manganite layer in ferroelectric/manganite heterostructures.[18] However, a clear IMT has not yet been achieved with the ferroelectric polarization control, most probably due to additional sample processes including nano-lithography, which induces more complications to the system. The absence of the experimental data hampered the understanding of the detailed mechanisms for ferroelectric field effect. It is worthy of note that there have been attempts to switch the polarization of the whole layer by using ambient gas[19] or mechanical pressure,[20] but the LSMO phase change using ferroelectric switching has been rarely explored in such cases.

In this letter, we show that IMT can be realized in LSMO by capping the layer with a ferroelectric PZT epitaxial layer. Correspondingly, the ferromagnetic phase transition temperature ($T_C$) and saturation magnetic moment ($M_s$) were also substantially tuned depending on the doping level, indicating a clear crossover across the phase boundary of LSMO due to ferroelectric field effect. Comparison with non-ferroelectric SrTiO$_3$ (STO) and LaAlO$_3$ (LAO) capping layers excluded other possible origins for the IMT in the LSMO layer, besides ferroelectricity.

A group of LSMO (with nominal $x = 0.2$) heterostructures were epitaxially fabricated by pulsed laser epitaxy on atomically smooth TiO$_2$-terminated (001) STO substrates. The thickness of the LSMO layer was varied as 5, 10 and 30 nm. We then deposited 10-nm-thick PZT, STO, and LAO ultrathin films as capping layers on top of the LSMO layers. The samples were fabricated at 625 °C in 100 mTorr of oxygen partial pressure. A KrF excimer laser ($\lambda = 248$ nm) with a laser fluence of ~0.2 J/cm$^2$ was used for ablating sintered PZT and LSMO targets and single crystal STO and LAO targets. Typical polarizations of our PZT thin films were ~80 $\mu$C/cm$^2$, with not much thickness dependent fluctuations. Details on the growth condition can be found elsewhere.[21]

Figure 1(a) shows a schematic diagram of our samples. The topography of a final PZT/LSMO heterostructure was observed by atomic force microscopy (AFM), showing an atomically-flat surface and well-defined terraces, as shown in Fig. 1(b). Note that the growth of atomically-flat, high quality heterostructures is a critical step towards obtaining the high polarization in PZT, required for effective ferroelectric switching with an ultrathin film.[21] Also note that the image is after positive and negative ferroelectric switching, which indicates that the surface topography is not influenced by the ferroelectric poling. The phase images of the PZT film with LSMO as a bottom electrode were measured by piezoresponse force microscopy (PFM) as shown in Fig. 1(c). The polarization was switched by a conductive AFM tip using +2.5 and −2.5 V. The PFM image of the PZT film reveals that it has a single as grown ferroelectric domain with an upward polarization direction and clear ferroelectric switching property. On the other hand, the PFM images from the LAO/LSMO (Fig. 1(d)) and STO/LSMO (data not shown) heterostructures, which we have grown to validate the ferroelectric field effect, indicate that both LAO and STO layers are indeed non-ferroelectric. Temperature ($T$)-dependent resistivity curves, $\rho(T)$, were recorded by a physical property measurement system (PPMS, Quantum Design Inc.). Ohmic indium contacts were ultrasonically soldered to the samples' corners in Van der Pauw geometry and, then, gold wires were bonded to the contacts as schematically shown in Fig. 1(a). $T$- and magnetic field ($H$)-dependent magnetization curves, $M(T)$ at 200 Oe and $M(H)$ at 10 K, were recorded using a 7 T superconducting quantum interference device (SQUID, Quantum Design Inc.) magnetometer.

When more holes are doped into LSMO (*i.e.*, $x$ is increased) near $x = 0.2$ across the phase boundary, we can expect two different phenomena as illustrated in Fig. 1(a): The system will undergo an IMT and the magnetic $T_C$ will increase with increasing $x$. Figure 2 demonstrates that by observing these two phenomena, we could successfully control the phase of LSMO not through chemically adding extra holes, but by electrostatically using a ferroelectric capping layer. As shown in Fig. 2(a), the bare LSMO film exhibited a highly insulating $\rho(T)$ behavior with a $T_C$ of ~200 K (marked with a triangle). On the other hand, the PZT/LSMO heterostructure exhibited a metallic $\rho(T)$ behavior over a wide range of $T$. Surprisingly, $\rho(T)$ of LSMO was decreased by at least four orders of magnitude at ~70 K by capping with the PZT layer. This ratio seems to be further increased at lower $T$ (the bare LSMO's resistivity at lower $T$ was above the instrumental limit.) The $T_C$ was also increased substantially to ~250 K, as expected from the field doping. The increase in ferromagnetic $T_C$ by the ferroelectric layer was confirmed by the $M(T)$ curves as shown in Fig. 2(b). Furthermore, an evident increase in $M_s$ in the LSMO layer, from 2.06 to 3.71 $\mu_B$/Mn, was observed with PZT capping, as shown in the $M(H)$ curves in the inset of Fig. 2(b).

It is obvious that the boundary condition of LSMO changes with ferroelectric capping, and this strongly affects the physical properties of the layer. In particular, the upward ferroelectric polarization of the as-grown PZT layer (see Fig. 1(c)) will attract more holes to the PZT/LSMO interface. Consequently, the carriers will be accumulated near the interface to screen the electric field, as schematically shown in Fig. 1(a). The increased hole density would increase the conductivity of the LSMO layer, eventually crossing the insulator-metal phase boundary. There can be an alternative picture, which does not necessarily contradict with the previous scenario. Note that both FM-I and FM-M phase domains coexist for LSMO near $x = $ ~0.2.[22,23] The hole accumulation due to the ferroelectric field effect would enhance the population and/or size of the FM-M phase domains, bringing the system above the percolation limit to exhibit the metallic behavior. The increased $T_C$ evidently supports this idea, because the increased magnetic double exchange interaction results in the increased fraction of FM-M phase domains.

In order to confirm that the large change in the physical properties of the LSMO layer is actually resulting from the ferroelectric layer, we deposited non-ferroelectric capping layers (STO and LAO) instead of PZT. As shown in Fig. 2, $\rho(T)$, $M(T)$, and $M(H)$ curves of LAO/LSMO are very similar to those of bare LSMO with no capping layer. This clearly illustrates that a non-ferroelectric layer does not change the electrostatic boundary condition in LSMO as compared to air (*i.e.* non-capping situation). On the other hand, the results on the STO/LSMO heterostructure deviated from the bare LSMO, although the change was weaker than that for PZT/LSMO. While more detailed study is required to exactly determine the role of STO capping layer, we suggest a charge transfer across the interface as one of the origins of the change in the physical properties in STO/LSMO. In particular, STO is known to easily supply oxygen to the LSMO surface that possibly has an oxygen deficient dead layer.[24,25] Thus, the changes in $\rho(T)$ and $T_C$ in STO/LSMO might be related with the fact that some of oxygen defects in LSMO were compensated by interfacing with STO. In addition, cation intermixing or doping to the interface between the LSMO and STO might also be possible. However, it unlikely based on our atomically flat topography and well-defined x-ray reflectivity fringes, indicating a rather sharp interface.

We further studied the dependence of the ferroelectric field effect on the thickness of LSMO. If the ferroelectric PZT layer has the largest influence near the PZT/LSMO interface, the effect will decrease as the LSMO thickness increases. Figure 3(a) shows $\rho(T)$ of 5, 10, and 30 nm of LSMO with and without a PZT capping layer. It certainly shows that the change coming from the PZT capping decreases as LSMO thickness increases. For bare LSMO, the resistivity was increased with decreasing thickness. When the thickness of the LSMO layer was smaller than 10 nm, *i.e.* for 5 nm LSMO, the film became highly insulating, suggesting that the critical thickness for the metallic LSMO is between 5 and 10 nm. A similar thickness dependent transport behavior has been reported for LSMO with $x = 0.3$ thin films,[3] where they suggested a loss of a conducting percolation path for LSMO layers thinner than ~3.2 nm. Those $x = 0.3$ LSMO films have more stable FM-M phases compared to the case of $x = 0.2$ reported here, which could explain the difference in the critical thicknesses. However, the same argument can be applied for the $x = 0.2$ LSMO films. When these LSMO films are capped with PZT, $\rho(T)$ decreases regardless of the LSMO thickness. However, the effect was the largest for the 5 nm LSMO film, the thinnest among studied here. While other contributions such as compensation of surface dead layer and/or prevention of carrier depletion due to capping layer might also play some role, it is evident that the ferroelectric field effect is the largest near the interface of PZT/LSMO interface with the thickness of several nanometers. Figure 3(b) shows the magnetic properties of the samples. It shows an increase in $T_C$ and $M_s$ with PZT capping, but again, the amount of change decreases with increasing LSMO thickness. We note that $M(T)$ curves for 30 nm LSMO with and without the PZT capping layer are different from others, *i.e.* it seems to show two different transitions with decreasing $T$. While it requires a detailed study on the magnetic property, we suggest that the strain relaxation leading to some lattice distortion in the thicker film might be attributed to such an exotic behavior.

Figure 4 summarizes the transport and magnetic properties of LSMO and PZT/LSMO heterostructures, as a function of the LSMO thickness. The resistivity ratio between PZT/LSMO heterostructure ($\rho_{PZT/LSMO}$) and LSMO film ($\rho_{LSMO}$) clearly increased as $T$ decreased. As expected, the resistivity ratio of the 5 nm-thick LSMO increased most rapidly. Surprisingly, at 50 K, we could obtain a resistivity ratio greater than five orders of magnitude, by extrapolating the $\rho(T)$ for the bare 5 nm LSMO. The extrapolation was done by fitting the curve using the Arrhenius function, which could fit other insulating curves (10 and 30 nm of LSMO at low temperature quite well. Figure 4 also shows the change in $T_C$ ($\Delta T_C$, $= T_{C,PZT/LSMO} - T_{C,LSMO}$) between PZT/LSMO and LSMO, for different LSMO thicknesses. For 5 nm-thick LSMO, the change in $T_C$ due to ferroelectric capping is increased by ~50 K. Finally, $M_s$ ratio at 10 K, at 0.2 T between PZT/LSMO and LSMO ($M_{s,PZT/LSMO}/M_{s,LSMO}$) is shown. The strongest effect of ferroelectric layer is again shown from the thinnest LSMO film, enhancing $M_s$ as large as twice. These results clearly indicate that the phase of LSMO is most susceptible to the ferroelectric polarization near the interface of PZT/LSMO.

In summary, substantial changes in transport and magnetic property of LSMO ultrathin films under the influence of ferroelectric polarization have been reported. As compared to the bare LSMO thin films, PZT/LSMO heterostructures showed a large modulation of resistivity, magnetic phase transition temperature, and saturation magnetization. Moreover, we have confirmed that an insulator-metal transition in manganites can be achieved when the manganite layer is sufficiently thin as the field effect is dominantly active near the interface, driven by ferroelectric polarization. Therefore, our unambiguous observations on the ferroelectric field effect control could aid understanding and controlling functional oxide heterostructures, opening a door to realizing oxide electronic devices such as ferroelectric field-effect transistors and non-volatile data storage.

We would like to thank Y. S. Kim for technical assistance for PFM measurements and E. Dagotto for helpful discussions. This work was supported by the U.S. Department of Energy, Basic Energy Sciences, Materials Sciences and Engineering Division.

**Figure captions**

Fig. 1. (a) Left and right images represent schematic drawings on a PZT/LSMO heterostructure and a LSMO thin film on STO, respectively. The electrodes for in-plane Van der Pauw transport measurements were contacted to the LSMO layer. The image in the middle shows a part of the phase diagram of $La_{1-x}Sr_xMnO_3$. (b) Topographic image of a PZT(10 nm)/LSMO(5 nm) heterostructure deposited on a (001) STO substrate, showing an atomically smooth surface with clear step-terrace structure. (c) PFM image of PZT on bottom electrode LSMO showing clear ferroelectric switching in the PZT layer. The contrast was produced by a conductive AFM tip under +2.5V and -2.5V, for downward (×) and upward (⊙) domain, respectively. Region I corresponds to as-grown background, and Regions II and III correspond to positively and negatively poled regions, respectively. (d) PFM image of a LAO on LSMO showing no piezoresponse. The scale bars on the images correspond to 400 nm.

Fig. 2. Transport and magnetic properties of ultrathin (5 nm) LSMO films with different capping layers of PZT, STO and LAO. $T_C$ is marked with a triangle. (a) Temperature-dependent resistivity for LSMO with 10 nm capping layers. (b) Temperature-dependent magnetization of LSMO and LSMO with 10 nm capping layers thereon measured at 200 Oe. Inset shows magnetic hysteresis loops for corresponding samples measured at 10 K.

Fig. 3. LSMO thickness dependent transport and magnetic properties of LSMO films with 10 nm-thick PZT capping layer. $T_C$ for LSMO/STO and PZT/LSMO/STO heterostructures are marked with empty and solid triangles, respectively. Temperature-dependent (a) resistivity and (b) magnetization for 5, 10, and 30 nm LSMO with and without PZT capping layers. Inset shows the magnetic hysteresis loops for corresponding samples measured at 10 K.

Fig. 4. $\rho(T)$ ratio, $\Delta T_C$, and $M_s$ ratio at 10 K at 0.2 T between LSMO and PZT/LSMO, for different LSMO thicknesses. The value for the open circle symbol in dark blue (50 K) was estimated from extrapolation by fitting the resistivity curves with the Arrhenius function. The thick solid lines are guide to the eye.

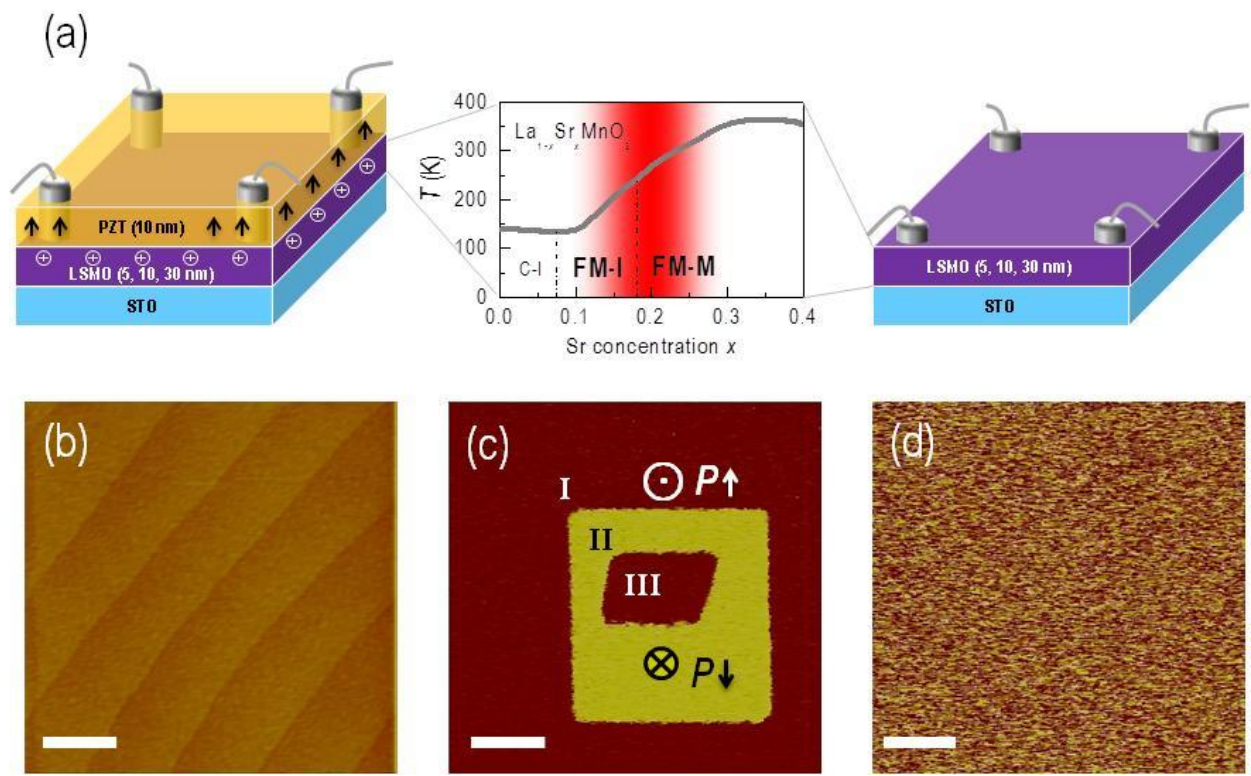

Figure 1.
Jiang et at.

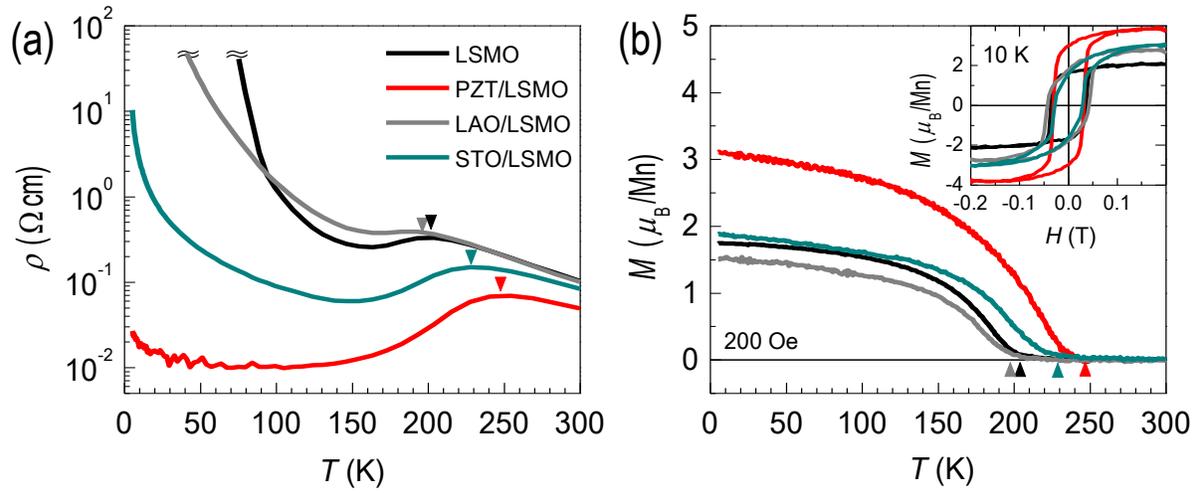

Figure 2.
Jiang et al.

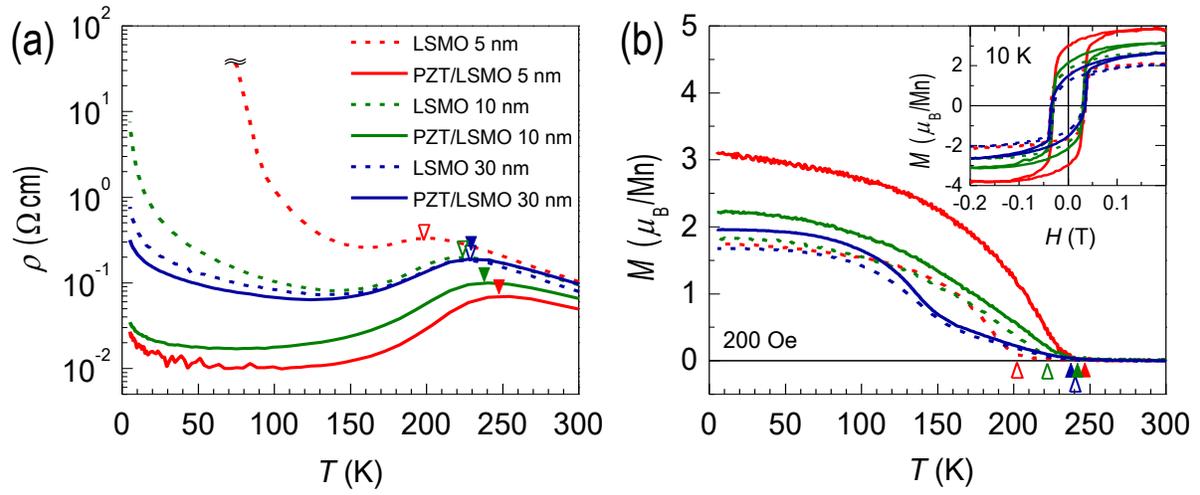

Figure 3.
Jiang et al.

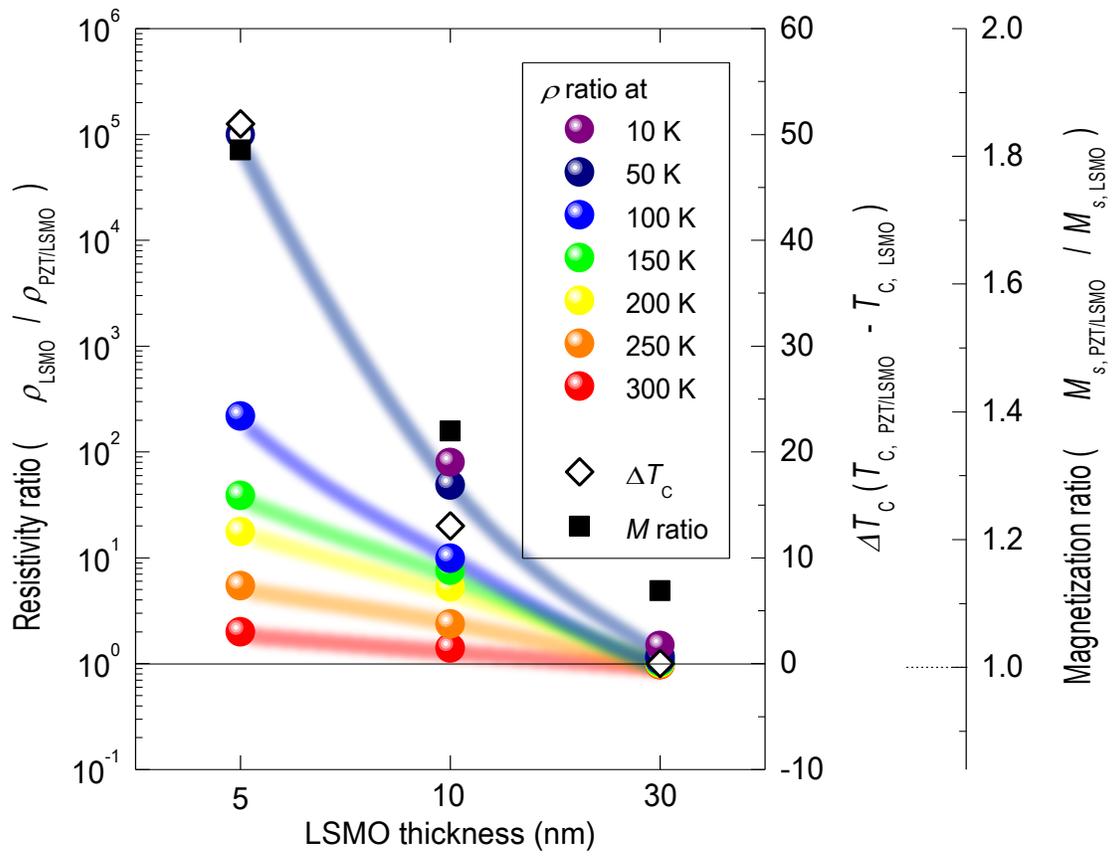

Figure 4.
Jiang *et al.*